\documentclass[pdflatex,sn-mathphys-num]{sn-jnl}

\usepackage{graphicx}%
\usepackage{multirow}%
\usepackage{amsmath,amssymb,amsfonts}%
\usepackage{amsthm}%
\usepackage{mathrsfs}%
\usepackage[title]{appendix}%
\usepackage{xcolor}%
\usepackage{textcomp}%
\usepackage{manyfoot}%
\usepackage{booktabs}%
\usepackage{algorithm}%
\usepackage{algorithmicx}%
\usepackage{algpseudocode}%
\usepackage{multirow}
\usepackage{listings}%
\usepackage{changes}
\usepackage{xcolor}
\usepackage{csquotes}
\definechangesauthor[name=Jörg, color=magenta]{Jhe}
\definechangesauthor[name=Bettina, color=brown]{BJ}
\definechangesauthor[name=Gerhard, color=orange]{GH}
\definechangesauthor[name=All, color=red]{ALL}
\usepackage{ulem}

\begin{document}

\title[Quantum Workshop for IT Professionals]{
Quantum Workshop for IT Professionals }


\author*[1,4]{\fnm{Bettina} \sur{Just}}\email{bettina.just@mni.thm.de}
\author[2]{\fnm{Jörg} \sur{Hettel}}\email{joerg.hettel@hs-kl.de}
\author[3]{\fnm{Gerhard} \sur{Hellstern}}\email{gerhard.hellstern@dhbw-stuttgart.de}

\equalcont{All authors contributed equally to this work.}

\affil[1]{\orgdiv{FB MNI}, \orgname{Technische Hochschule Mittelhessen (THM)}, \city{Gießen}, \country{Germany}}

\affil[2]{\orgdiv{Department of Computer Sciences and Microsystems Technology}, \orgname{University of Applied Sciences Kaiserslautern}, \orgaddress{\country{Germany}}}

\affil[3]{\orgdiv{Center of Finance}, \orgname{Cooperative State University of Baden-Württemberg (DHBW)}, \orgaddress{\city{Stuttgart}, \country{Germany}}}

\affil[4]{\orgdiv{Competence Center for Quantum Computing}, \orgname{TransMIT}, \city{Gießen}, \country{Germany}}

\abstract{
Quantum computing is gaining strategic relevance beyond research-driven industries. However, many companies lack the expertise to evaluate its potential for real-world applications. Traditional training formats often focus on physical principles without demonstrating practical relevance for their Business Processes, which limits success.

This paper presents a user-centered workshop concept tailored to IT professionals without prior quantum knowledge. Using a business simulation game set in a fictitious company, participants explore quantum technologies through relatable, application-driven scenarios. The flexible design allows customization for different organizational contexts.

Evaluation results from a one-day implementation at the IT-Tage 2024 indicate clear learning progress and increased awareness of practical use cases. The approach effectively bridges the gap between complex quantum concepts and industry-specific application needs.
}

\keywords{Quantum computing, quantum information, quantum algorithms, education}



\maketitle

\section{Introduction}\label{introduction}
Second-generation quantum technologies are becoming more and more important \cite{Dowling_2003}. 
Quantum information and quantum computing are increasingly capturing the interest of companies whose business models do not primarily revolve around fundamental quantum technologies.
These companies now face the major challenge of embracing the new opportunities offered by these technologies. The ongoing digital transformation must be expanded to include a quantum transformation in order to succeed in the market. However, the companies generally have no expertise in these new areas to correctly assess the future potential inherent in quantum technologies. 
Many companies lack the experience to make strategic decisions about how they can open up new business areas or improve existing processes with the new quantum technologies. Unfortunately, this means that future competitive advantages are not realised.

Customised training concepts must be developed to support companies in the strategic evaluation of the new possibilities for the use of quantum information and quantum computing in particular.
The target group for the training concepts must be the company's employees and managers, as it is precisely this target group that has the necessary and valuable knowledge from the business area. Unfortunately, this target group regularly lacks the necessary application knowledge from the quantum world. The questions that now arise are: How can a company's employees and management be given an introduction to quantum information and quantum computing in a simple and suitable way? And: How can a company utilise the future potential of these technologies to gain a competitive edge?

Companies interested in quantum technologies are typically offered introductory training sessions that provide an overview of the underlying physical principles. These sessions often emphasize foundational concepts in physics, including, for example, the Schrödinger equation. However, a connection between these physical fundamentals and potential industry-specific applications is frequently lacking. But it is precisely these questions about the profitable use of quantum technologies that are at the forefront of companies' minds. As a result, the perceived relevance for companies remains limited, and the topic of quantum technologies is often not pursued further, as concrete application potentials tailored to the company’s context are neither demonstrated nor identified.

To address these issues the authors have developed an innovative workshop concept that is tailored precisely to this target group. The workshop concept was developed by the authors as an accompanying event to a series of quantum MOOCs and was held for the first time in this context \cite{Hellstern_2024}. Since its inception, the workshop concept has undergone further development. This article discusses its current state and the insights derived from its implementation. As far as the authors are aware, there is as yet no comparable approach to introduce the possibilities of quantum computing to companies. The workshop takes the form of a business simulation game in which a fictitious company is made “quantum-ready” for the future. 
A business simulation game is used as a model-based learning and training tool that realistically depicts business management contexts. The focus is on economic business processes such as production, finance, marketing, and corporate strategies. The relevance of the content for the economy is a priority, as this is the deciding factor for the adoption of new technologies \cite{faisal2022business}. In our case each participant takes on a role in a fictional company. The participants work together and do not compete with each other. The actual business simulation gam is carried out without computer support, using only pens and paper or tactile materials. Depending on the interests of the participants, computers may be used during the workshop, e.g., for small programming projects. The workshop concept and the main subjects are kept open so that it can be adapted to different specific target groups and companies.

The latest iteration of the workshop took place during the IT-Tage 2024 in Frankfurt am Main. The IT-Tage are one of the largest user conferences in the IT industry in Germany\footnote{https://www.ittage.informatik-aktuell.de/programm/2024/workshop-quantencomputing-in-der-praxis.html}. At the IT-Tage, the workshop was compressed into one day for organisational reasons. On other occasions, the workshop was held over two or three days.
In order to further improve the concept and benefits, we began evaluating the workshops. At the IT-Tage, a survey was conducted among participants before and after the workshop.
The following research questions are of particular interest in order to evaluate the success of the workshop and to further improve the concept:

\begin{description}
\item[\bf RQ 1:] 
Individual learning progress after the seminar: How do the participants evaluate their individual learning progress in the basics and possible applications of quantum computing after completing the one-day seminar?

\item[\bf RQ 2:]
Relevance of specific quantum computing applications: Which specific applications of quantum computing do the participants identify as particularly relevant for their professional practice?

\item[\bf RQ 3:] 
 Challenges of implementation in IT systems: What specific challenges do participants identify when implementing quantum computing technologies in existing IT systems and processes?
\end{description}

This paper is organized as follows. Section~\ref{literature_review} provides an overview of relevant publications on quantum computing education related to this paper.  Section~\ref{design} presents the didactic design and the topics of the workshop under consideration in this paper.
The open concept of the workshop requires the leaders to have a broad range of expertise. This necessary knowledge is also categorized with respect to updated European Competence Framework for Quantum Technologies version 2.5 (2024) \cite{Greinert_2024a}. \\
Section~\ref{results} analyses the data collected at the beginning and at the end of the workshop with regard to the three research questions mentioned above.
In Sect.~\ref{conclusion}, based on the results of Sect.~\ref{results}, we draw conclusions concerning the research questions, the lessons learned, and point out topics for further research and improvements to the workshop concept.

\section{Literature review}
\label{literature_review}
The rapid development of quantum technology has created an urgent need for skilled professionals, necessitating a comprehensive transformation of the educational and training landscape. This literature review examines the necessity of quantum workforce training, the usage of competence frameworks, and concrete measures to promote education in this field.

\subsection*{Necessity of Quantum Workforce Training}

Quantum technology, with its potential applications in various fields such as quantum computing, quantum communication, and quantum sensing, is a rapidly growing field that attracts significant investments from governments and companies worldwide \cite{Aparicio-Morales_2024}. These investments aim to increase global competitiveness and position respective countries as key players in this area. The commercialization of quantum technologies leads to a substantial increase in demand for skilled professionals, resulting in a shortage of talent in this field \cite{Kaur_2022}.

A significant aspect is that the current quantum workforce primarily originates from the academic sector, particularly from physics PhDs, and there is a lack of professionals at the non-doctoral level \cite{Gerke_2022}. However, the demand for quantum experts is growing not only in research but also in various industries, requiring other professionals such as engineers, software developers, product designers, and strategists \cite{Greinert_2024}. This development highlights the need for a broader educational strategy that not only targets highly specialized quantum physicists but also introduces professionals from other fields to quantum technology.

The growing gap between industry demand and existing qualifications requires targeted educational initiatives \cite{Greinert_2024}. It is obvious that academic degrees alone are not sufficient to meet job market demands. Practical experience through internships, projects, and other technical skills are as important as theoretical knowledge \cite{Tenjo-Patino_2024}. Adapting curricula to industry needs is therefore an essential part of efforts to develop a quantum-competent workforce \cite{Purohit_2024}.

Moreover, there is the necessity of "Quantum Literacy" for the general population. As quantum technologies penetrate everyday life, it is important that people outside the quantum industry also understand the fundamentals \cite{Greinert_2024}. This development requires the teaching of quantum concepts in schools and to the public.

\subsection*{Quantum Competence Framework for the Workforce}

In order to structure and standardize education in quantum technologies, the European Competence Framework for Quantum Technologies (CFQT) has been developed. It serves as a reference framework for planning, comparing, and evaluating educational activities, qualifications, and job requirements. The CFQT is an important tool for standardizing quantum education and ensuring comparability of qualifications and training programs in Europe \cite{Greinert_2023}.

The CFQT has been developed from studies on competencies, requirements, and predictions for the future quantum workforce \cite{Gerke_2022}. It serves as a common language for quantum education and is used in projects such as DigiQ and QTIndu \cite{Kaur_2022}. The current version of the CFQT includes eight domains with 42 subdomains covering various topics. The latest version, 2.5, also includes a profiency triangle and qualification profiles that provide more detailed descriptions of competence levels and typical in the quantum industry \cite{Greinert_2024}. The proficiency triangle defines and visualizes six proficiency levels in the three proficiency areas \enquote{quantum concepts}, \enquote{quantum hardware and software engineering} and \enquote{quantum technology applications and strategy}, see also figure \ref{fig:CompetenceTrainers}. The proficiency is very useful to visualize quantum competences.

In addition to the CFQT, there are other initiatives to create frameworks for quantum education and workforce development. These include, for example, the national strategy plans of various countries such as the USA, which aim to promote collaboration between educational institutions, industry, and government \cite{Aiello_2021}. Industry consortia like the Quantum Economic Development Consortium (QED-C) in the USA are also working on defining qualification requirements and promoting collaboration between industry and academic institutions \cite{Bonilla-Licea_2024}. These frameworks aim to create a unified understanding of the required competencies and qualifications to accelerate the development of the quantum workforce.

\subsection*{Concrete Measures for Quantum Workforce Training}

Various concrete measures are being taken to train the quantum workforce. These measures include both traditional educational offerings and innovative and informal approaches:

\begin{itemize}

\item \textbf{Promotion of Quantum Education in Schools:} Initiatives are being undertaken to promote quantum education in schools and spark interest in quantum technologies among young people. Providing teaching materials and training teachers are important components of these efforts \cite{Gerke_2022}.

\item \textbf{Adapting Higher Education:} Universities and colleges are developing new courses of study specifically focused on quantum sciences and technologies \cite{Greinert_2024}. These programs include both theoretical and practical aspects and are tailored to industry requirements \cite{Purohit_2024}.

\item \textbf{Development of Master's and PhD Programs:} Specialized master’s and PhD programs are being established to train highly qualified professionals for research and industry. These programs often include interdisciplinary approaches and collaboration with national laboratories and leading universities \cite{Greinert_2024}.

\item \textbf{Online Educational Offerings:} There are a variety of online courses and platforms that facilitate access to quantum education. These offerings are often flexible and allow learners to study at their own pace \cite{Kaur_2022,Hellstern_2024}.

\item \textbf{Community Building and Networking Events:} Various community-building initiatives and networking events are organized to raise awareness of quantum technologies and connect students and professionals \cite{Venegas‐Gomez_2020}. These activities include workshops, mentoring programs, and conferences that promote the exchange of knowledge and experiences \cite{Greinert_2023}. In addition, in the last years a couple of non-profit organiztions like Unitary Foundation (https://unitary.foundation/), Quantum AI Foundation (https://www.qaif.org/) started to organize hackathons and training initiatives which partly also are targeted to workforce training.

\item \textbf{Experience-Oriented Learning Methods:} Innovative learning methods, such as the ENSAR (Experience-Name-Speak-Apply-Repeat) concept and the use of tools like the Qureka! Box, are used to convey complex quantum concepts in an accessible way \cite{Purohit_2024}. These methods emphasize practical application of what is learned and foster understanding through active participation.

\item \textbf{Professional Development Programs:} Special training programs are being developed for professionals who want to further qualify in quantum technologies \cite{Aiello_2021}. These programs are aimed at engineers, software developers, and other professionals who want to become familiar with quantum concepts and applications \cite{Chrisochoides_2024}.

\end{itemize}

These concrete measures show that various actors worldwide are working on developing a qualified quantum workforce. However, the challenges are diverse and require continuous adaptation and further development of educational strategies. 
Apart from the literature about Professional Development Programs cited above, we observed a shortage of studies about this topic. This indicates a research gap which have to be filled to support efforts for customized training programs in the professional sector.

\section{Design of the workshop}\label{design}
Imparting knowledge through user-centered workshops has the advantage that the needs and expectations of the participants can be specifically addressed.
In this section, the didactic and content design of the workshop is explained in more detail.

\subsection{Workshop concept and structure}
Successful learning requires a response to the needs  of the participants. It is essential to understand their motivation and prior knowledge . Moreover, throughout the workshop, small learning checks are carried out so that the participants can experience their growth in competence  \cite{Hattie_2017_mindframes}. \\
The workshop starts with an opening using a so called "living statistics". 
Participants divide themselves into groups according to their current position in their profession. This allows participants with similar tasks to be quickly identified. Further grouping can be based on areas of interest, prior knowledge, or company size. The fact that participants come together several times according to different criteria creates interaction between people with similar interests and backgrounds at an early stage. 
We observed that this was effective
for making the participants feel welcome and taken seriously, to activate them and to create a positive learning environment. 
As a next step, before participants begin identifying potential applications and developing solution or product concepts, a series of keynote presentations are delivered by the trainers. These presentations are spontaneously adapted to reflect the specific interests and backgrounds of the participants, which were determined by the “living statistics” and accompanying discussions. One focus here is always on what quantum technology could achieve today and in the future.
It is important that the trainers are flexible enough to respond to questions, to adjust the focus of the presentation and to explore the desired topics in greater depth.

Common topics covered include foundational principles of quantum computing, quantum cryptography and security, the quantum internet and various protocols, as well as the design and programming of quantum algorithms.

Following the presentations, participants will be introduced to a fictional company that is now to be made “quantum-ready.” This fictional company has a standard corporate structure with the following departments: management, accounting, controlling, human resources, production units, research and development, IT services, etc. Since participants usually come from companies, they are familiar with these structures and can therefore immediately identify with a role in this company.
After that, participants are divided into working groups typically with 5 to 8 members. Each group selects a specific problem scenario from the context of the fictitious company and explores it in greater depth with guidance from the trainers. 

Typical challenges include, for example, the question of how quantum-based optimization algorithms could help reduce material consumption in production, or how quantum simulations could be used in materials research. Other topics include how quantum-based secure copy protection can be implemented and rolled out to combat product piracy. Can new business processes be derived from quantum protocols such as certified deletion or delegated computation? How should IT security position itself, and what new opportunities arise from the use of quantum networks? At the management level, discussions are underway on how to initiate a quantum transformation and what actions need to be taken, including knowledge and skills development for personnel.

To date, three main domains have consistently been in the center of attention in the workshop. These are IT operations and security, software development and programming, and executive or managerial leadership. 

Participants are encouraged to engage with the topic that best suits their interests. Participants often bring perspectives from their companies to the table, thereby enriching the discussion. Should a participant feel that their selected group does not align with their preferences or expertise, they are free to switch groups at any time.

At the end of the workshop, the working groups present the ideas and solutions they have developed to each other. For the conclusion of the workshop, it is crucial to derive concrete recommendations for action and follow-up activities to ensure that the workshop’s objectives continue to be pursued beyond the duration of the event.

During the entire workshop, special attention is paid to each individual participant and to the group dynamics within the learning group. Participants are encouraged to contribute their own topics and questions, which in turn can foster intrinsic motivation and is able to support long-term learning outcomes. Managing the group dynamics in the workshop is particularly challenging because the participants are diverse not only in terms of their background and interests, but also in terms of their professional qualifications in general and their prior knowledge of quantum computing in particular \cite{Koenig_2020_Gruppen}.

The workshop concept includes, if necessary, the flexible use of suitable methods for motivation, learning control, activation of the participants or for exchange among the participants \cite{Aerssen_2018_Innovation}.

\subsection{Motivation for using a business simulation game}
The use of a business simulation game puts learning and the activity of the participants to the focus and significantly supports a shift from teaching to learning \cite{Herz_2000_Simulation,Wildt_2003_shift}. The use of a simulation creates a realistic framework so that the participants can experiment with ideas and actions in a protected environment, and the time frame is adhered to. In addition, the leaders of the workshop can intervene effectively to promote the achievement of learning objectives for each single participant \cite{Herz_2000_Simulation}.
 This approach enables participants to integrate real-world challenges and requirements from their everyday work into the simulation, thereby enhancing both relevance and engagement.

Notably, the simulation also encourages participants to explore perspectives beyond their current real-life positions, often opting for roles outside their usual scope of responsibility.

\subsection{Requirements for workshop leaders}
\begin{figure}
    \centering
    \includegraphics[width=1.0\linewidth]{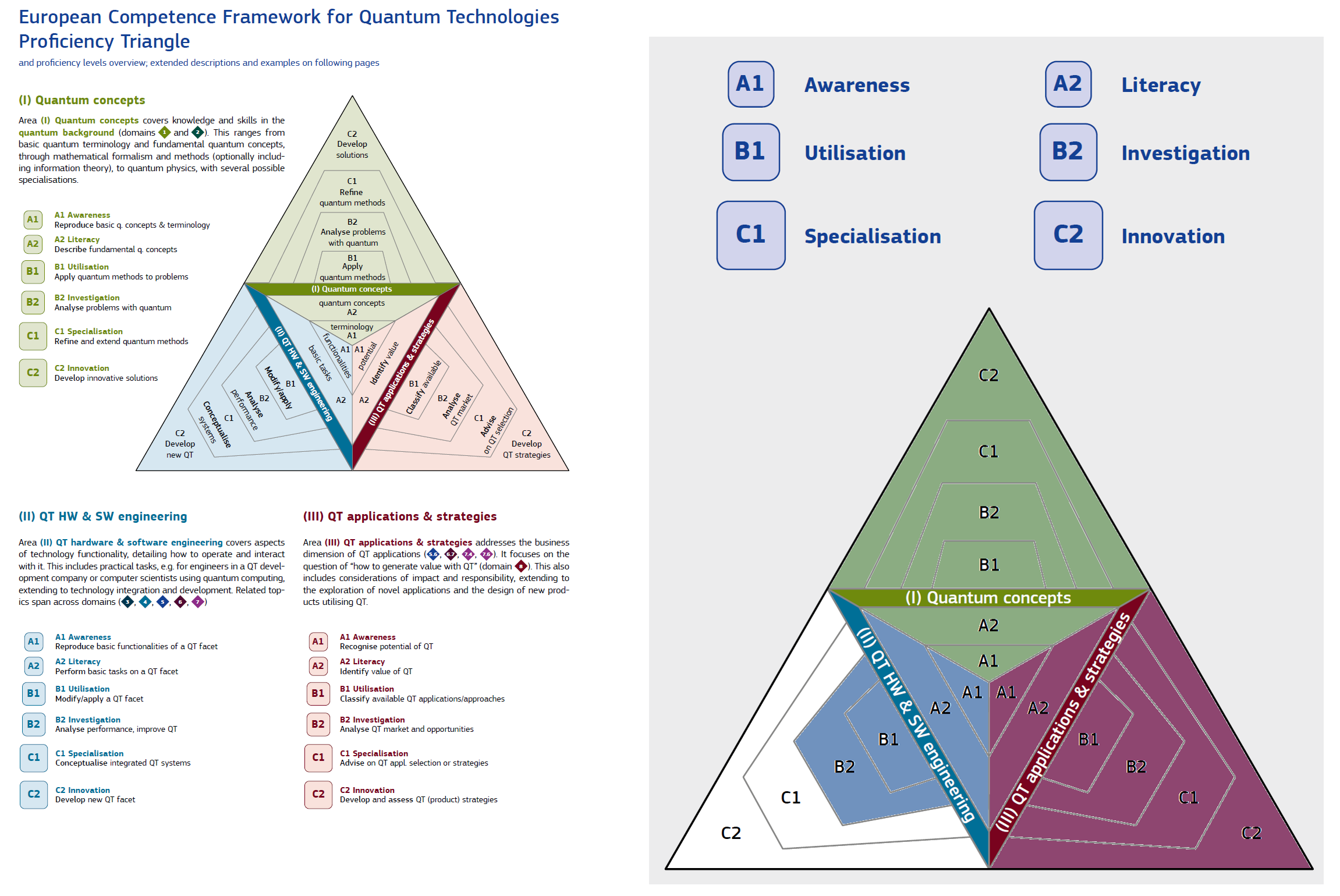}
    \caption{Left side: The European Competence Framework Proficiency Triangle \cite{Greinert_2024a} with the 6 proficiency levels in the three proficiency areas \enquote{quantum concepts}, \enquote{quantum hardware and software engineering} and \enquote{quantum technology applications and strategy}. The right side shows the skills of the workshop's trainer team by coloring the corresponding areas in the proficiency triangle. It turns out that the highest competence level C2 is required in the areas \enquote{quantum concepts} and \enquote{quantum technology applications and strategy}, and the third highest level B2 is required for the area \enquote{quantum hardware and software engineering}.}
    \label{fig:CompetenceTrainers}
\end{figure}

It is important to emphasize that the workshop does not follow a rigid, predefined curriculum. Instead, it is consistently oriented toward the specific interests and needs of the individual participants. 
This participant-centered approach requires a high degree of flexibility and a broad, interdisciplinary knowledge base from the seminar leaders – not only in the field of quantum technologies, but also with regard to teaching concepts as well as to real business processes.
The technical requirements for the trainers regarding quantum technology knowledge, derived from previous experience, are classified in the competency framework in figure \ref{fig:CompetenceTrainers}. Typically, the workshop should be accompanied by several trainers, with each trainer covering different specialist areas.

In addition, workshop leaders must demonstrate a high degree of creativity and spontaneity as well as the ability to effectively manage group dynamics.

\section{Evaluation of the data collected during a workshop in 2024}\label{results}
Since the workshop concept described above proved successful in its initial implementations, we have begun to systematically evaluate the concept in order to improve and expand it.
During the IT-Tage 2024 the workshopn described above was offered to interested participants.
The workshop was open to regular conference attendees; however, participation was subject to prior registration due to a restricted number of available places, which were restricted to 30. In the end, 24 participants joined the workshop and all of them took part in an online survey at the beginning of the workshop:
We asked for demographic information about the participants, their prior knowledge, their expectations/wishes, and what challenges they see in relation to the topic. At the end of the workshop
there were still 20 participants in the final online survey because some of them had to leave early. 
The second survey asked participants to assess their learning progress, what specific knowledge/skills they gained from the workshop, how relevant it was to their everyday working lives, including the associated challenges, and whether their expectations of the workshop and the trainers had been met.
An anonymized identifier made it possible to combine the answers of the participants in the two surveys before and after the workshop. Therefore, it is possible to track, e.g. the perception of knowledge gain.
Although the data collected so far are not yet statistically relevant, general trends can be identified and are therefore already meaningful to a certain extent.

\subsection{Information about the participants of the workshop}
\begin{figure}
   \begin{minipage}[b]{.5\linewidth} 
      \includegraphics[width=\linewidth, height=4cm]{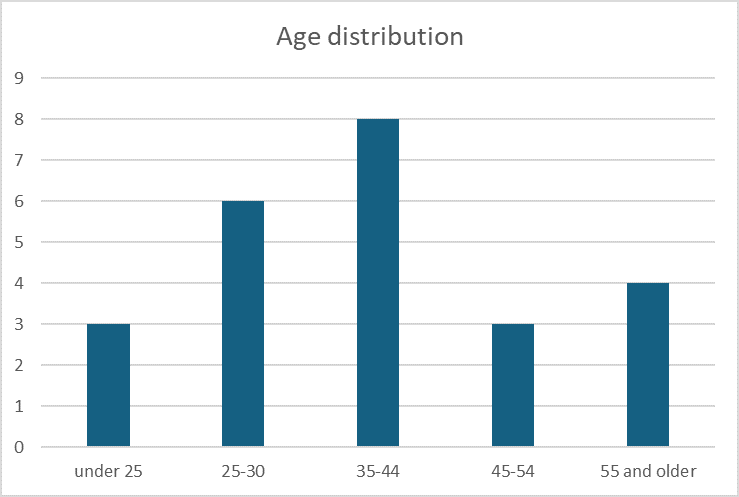}
      \caption{Age distribution of the participants}\label{Fig_1}
   \end{minipage}
   \hspace{.01\linewidth}
   \begin{minipage}[b]{.5\linewidth} 
      \includegraphics[width=\linewidth, height=4cm]{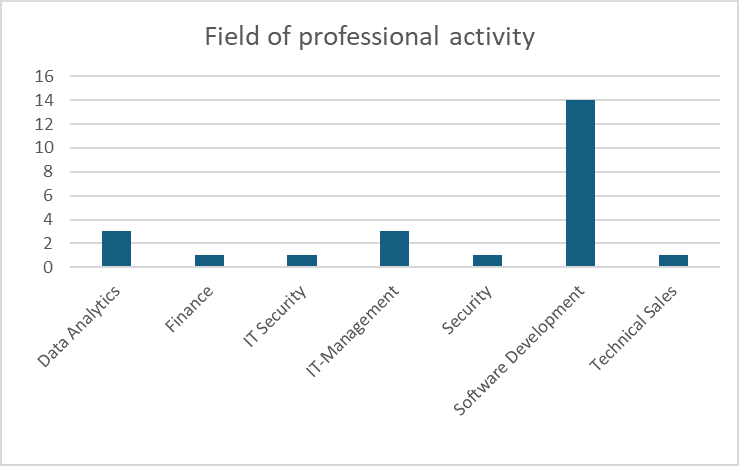}
      \caption{Field of professional activity}\label{Fig_2}
   \end{minipage}
\end{figure}

\begin{figure}
   \begin{minipage}[b]{.5\linewidth} 
      \includegraphics[width=\linewidth, height=4cm]{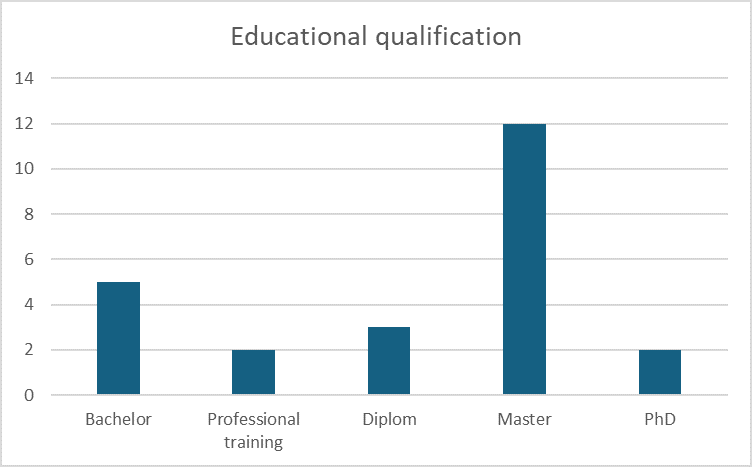}
      \caption{Educational qualifications}\label{Fig_3}
   \end{minipage}
   \hspace{.01\linewidth}
   \begin{minipage}[b]{.5\linewidth} 
      \includegraphics[width=\linewidth, height=4cm]{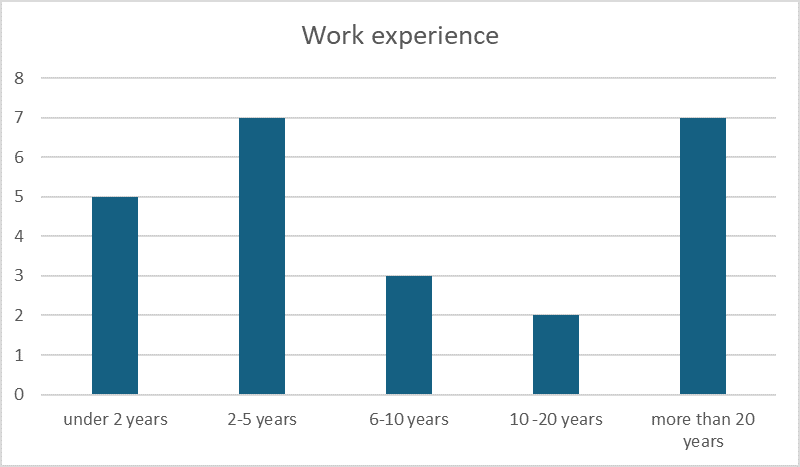}
      \caption{Work experience of the participants}\label{Fig_4}
   \end{minipage}
\end{figure}

This information has been collected at the beginning of the workshop.
In contrast to a typical course at a university, the composition of the participants was significantly different. As can be seen in Fig. (\ref{Fig_1}) the majority of participants were older than 35. Naturally, the predominant occupations are in the IT sector, such as software development or data analytics, see Fig. (\ref{Fig_2}). A few participants assigned themselves to IT management or finance. 
The majority of those present had a Master's degree or a “Diplom”\footnote{The German Diplom is a pre-Bologna university degree comparable in level to a master’s degree.} as their highest educational qualification - the rest had a Bachelor's degree, see Fig. (\ref{Fig_3}).

The extensive professional experience acquired by the participants is shown in Fig. (\ref{Fig_4}): Around half (12 participants) had more than 6 years of professional experience; 7 participants even had more than 20 years.   

\begin{figure}
   \begin{minipage}[b]{.5\linewidth} 
      \includegraphics[width=\linewidth, height=4cm]{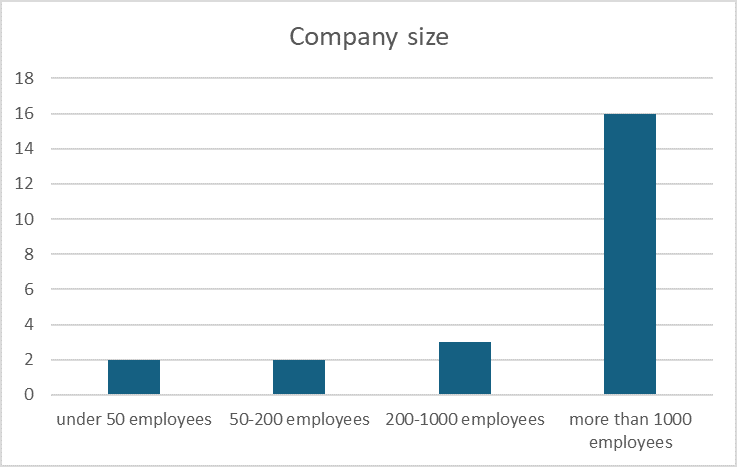}
      \caption{Company Size of the participants}\label{Fig_5}
   \end{minipage}
   \hspace{.01\linewidth}
\end{figure}

Among employers, large companies with more than 1000 employees predominate, followed by companies with 200-1000 employees, see Fig. (\ref{Fig_5}).

In summary, this gives a picture of the workshop participants who are in the middle of their careers and are either specialists or decision-makers with responsibility in their companies. Apart from a few consultants, it was mainly employees of large companies who were present. This means that the participants' solid prior knowledge of IT, including processes and procedures, is built upon. 

\subsection{Results for the research questions}

In the next three subsections we discuss our results to answer the research questions be proposed above.

\subsubsection{RQ1: How do the participants evaluate their individual learning progress in the basics and possible applications of quantum computing after completing the one-day seminar?}

To answer RQ1, we asked the participants at the beginning of the workshop how they assessed their previous knowledge.
Nobody assessed its knowledge as "adavanced"; most of them stated "no knowledge" or "basic knowledge". We deliberately left open the question of how “knowledge” could be acquired in order to include all potential avenues.

\begin{figure}
   \begin{minipage}[b]{.5\linewidth} 
      \includegraphics[width=\linewidth, height=4cm]{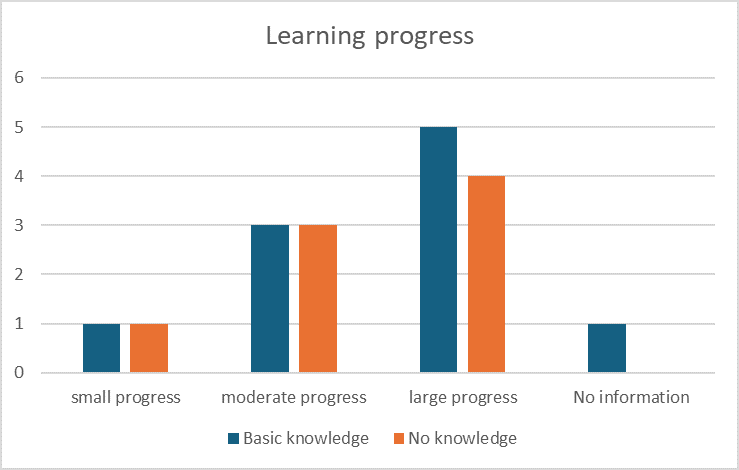}
      \caption{Perception of learning progress}\label{Fig_6}
   \end{minipage}
   \hspace{.01\linewidth}
   \begin{minipage}[b]{.5\linewidth} 
      \includegraphics[width=\linewidth, height=4cm]{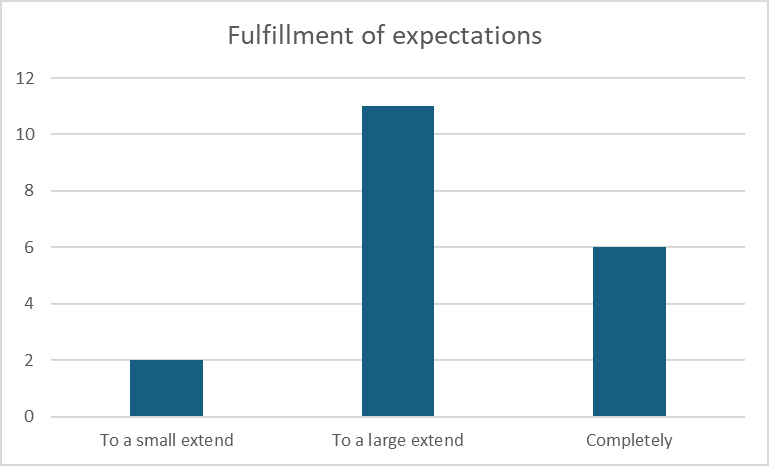}
      \caption{Fulfillment of expectations}\label{Fig_7}
   \end{minipage}
\end{figure}

Secondly, before the end of the workshop, we asked participants how they assessed their learning progress: Here, only “small progress”, “moderate progess” and “large progress” were selected (in addition to one participant with no information). Fig. (\ref{Fig_6}) shows how these categories - separated according to previous knowledge - are distributed among the participants. The answer “large progress” predominates, followed by “moderate progress”, regardless of previous knowledge.

The success of the workshop concept - imparting quantum know-how to participants from professional life - is also reflected in the extent to which the expectations of the participants were met, see Fig. (\ref{Fig_7}): For the vast majority, the expectations were met either to a large extent or even completely.

\subsubsection{RQ2: Which specific applications of quantum computing do the participants identify as particularly relevant for their professional practice?}

\begin{figure}
   \begin{minipage}[b]{.5\linewidth} 
      \includegraphics[width=\linewidth, height=4cm]{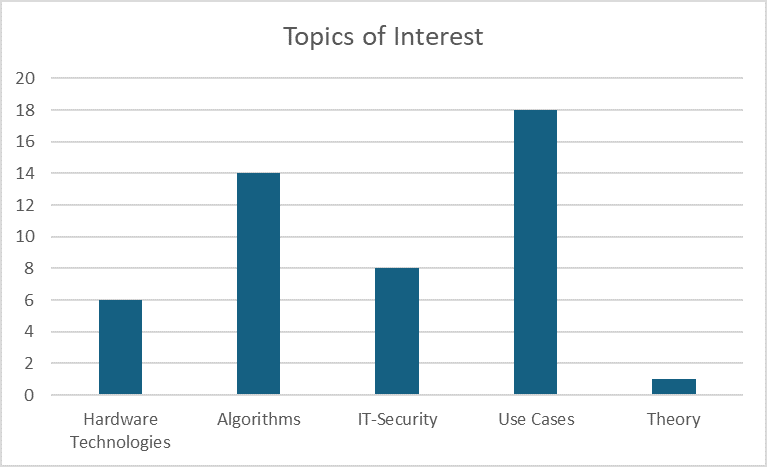}
      \caption{Topics of interest}\label{Fig_8}
   \end{minipage}
   \hspace{.01\linewidth}
   \begin{minipage}[b]{.5\linewidth} 
      \includegraphics[width=\linewidth, height=4cm]{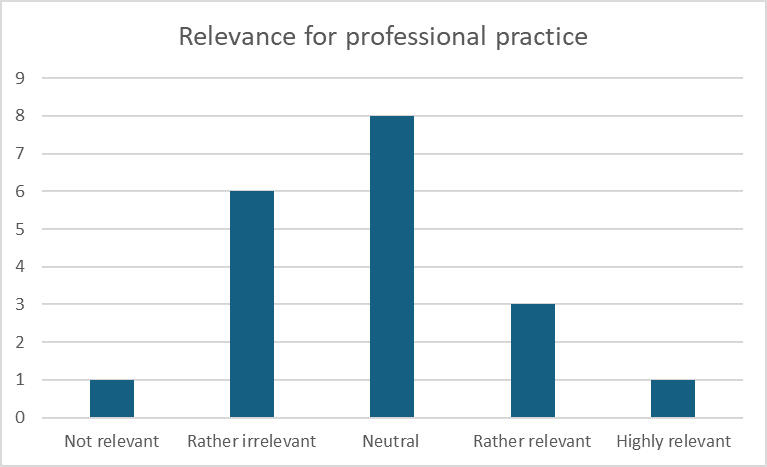}
      \caption{Relevance for professional practice in the day-to-day business of the participants.}\label{Fig_9}
   \end{minipage}
\end{figure}

When asked which quantum topics are of particular interest to the participants, the topic of “use cases” predominated, followed by “algorithms” and “IT security”, see Fig. (\ref{Fig_8}). However, the majority did not yet consider these quantum topics to be particularly relevant yet, see Fig. (\ref{Fig_9}).

In order to determine which topics may be relevant in the future of the participants we asked to mention the topics and how certain the participants are in their assessment. It was possible to mention several topics. The results are shown in Table (\ref{Tab1}) where the topics are sorted in accordance to certainty of the assessment.
Algorithms for various applications, as well as topics related to encryption, are attributed the highest relevance.

\subsubsection{RQ3: What specific challenges do participants identify when implementing quantum computing technologies in existing IT systems and processes?}

\begin{figure}
   \begin{minipage}[b]{.5\linewidth} 
      \includegraphics[width=\linewidth, height=4cm]{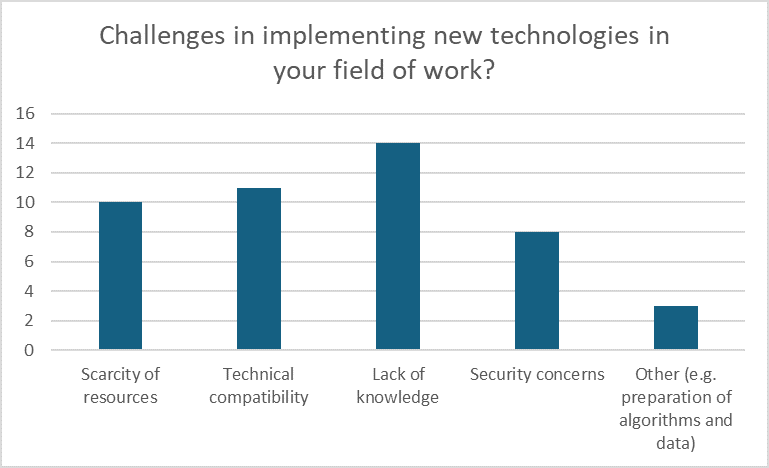}
      \caption{Challenges in implementing new technologies}\label{Fig_10}
   \end{minipage}
   \hspace{.01\linewidth}
\end{figure}

When asked about the challenges and difficulties in introducing quantum technologies in the participants' respective fields of work, the responses 'Lack of Knowledge' and 'Technical Complexity' were predominant, followed by 'Scarcity of Resources', see Fig. (\ref{Fig_10}). In the open-ended question (see Table \ref{Tab2}), a similar picture emerges: here too, aspects related to knowledge, the complexity of the topic, and unclear economic potential are 
apparent.

\begin{table}[]
\begin{tabular}{|c|c|}
\hline
Relevance of topics & Assessement \\
\hline 
Algorithms & \multirow{7}{*}{Rather certain} \\

Algorithm and basics & \\
Encryption & \\
Quantum programming, quantum machine learning & \\
(Post-quantum) cryptographic protocols & \\
Optimization & \\
Application to optimization problems & \\
\hline 
Optimization problems & \multirow{8}{*}{Neutral} \\
QKD & \\
Optimization, Security & \\
Encryption & \\
Making encryption quantum secure & \\
Encryption & \\
Key exchange in communication & \\
\hline 
Security and algorithms & Rather uncertain \\
\hline
\end{tabular}
\caption{Relevance of workshop topics for professional practice.}
\label{Tab1}
\end{table}

\begin{table}[]
\begin{tabular}{|l|}
\hline
"Currently, implementation is not an issue (other areas are dealing with the topic)"
\\
"For me, it's just about basic knowledge in case quantum computing is approached by customers." \\
\hline 
"Device / Hardware" \\
\hline 
"Lack of knowledge and resources" \\
\hline 
"Finding use cases" \\
\hline 
"High host of entry, little urgency because the cloud has good alternatives for everything" \\
\hline 
"Availability of reliable hardware" \\
\hline 
"Cost-benefit assessment from an entrepreneurial perspective. QC has potential
\\
but the question of whether or not people or employees are given time for the topic is very \\
dependent on the company. In principle, it is still necessary to identify which 
\\
consulting services can be provided as a result." \\
\hline
"Assessment of which problems can be mapped as optimization problems. \\
Further training of employees" \\
\hline 
"Use case too complex \\
Lack of investment from government and industry in Germany" \\
\hline 
"Outside of a simulation using Python, not feasible for me so far or 
\\
the technical requirements are not met" \\
\hline 
"Relevance in data processing" \\

\hline
\end{tabular}
\caption{Challenges in the implementation of quantum computing in the field of work. Original quotes of the participants translated from German to English. Not all of the participants answered this question.} 
\label{Tab2}
\end{table}

\section{Conclusions and topics for further research}\label{conclusion}
In the last section, the results of the data collection were presented.
This provides answers to the research questions formulated at the beginning:

\begin{description}
    \item[RQ1:] How do the participants evaluate their individual learning progress in the basics and possible applications of quantum computing after completing the one-day seminar?
\end{description}
The evaluations show that the majority of participants had the subjective impression that they had learned something important. In particular, even those with no prior knowledge were able to gain a lot from the workshop. This is also supported by the fact that the expectations of the vast majority of participants were met.
The data thus support the hypothesis that the workshop concept used provides a good introduction to this difficult subject matter. Participants feel really involved. Nevertheless, it should be noted that all statements made so far are based on self-assessment.

\begin{description}
    \item[RQ2:] Which specific applications of quantum computing do the participants identify as particularly relevant for their professional practice?
\end{description}
The application areas most frequently mentioned here, algorithms and security, were no surprise. 
This was partly due to the participants, who all came from the IT sector, and partly to the fact that these areas are also regularly featured in the general press.
It's remarkable that many participants are interested in use cases. This raises the question of specific areas of application. This is precisely where the strength of the workshop concept becomes apparent. It allows participants' needs and interests to be addressed much more effectively than with a training concept that emphasizes knowledge transfer.

\begin{description}
    \item[RQ3:] What specific challenges do participants identify when implementing quantum computing technologies in existing IT systems and processes?
\end{description}
Most participants were unable on this question to give an assessment based on their level of knowledge, or were uncertain. This uncertainty exists not only among workshop participants, but also in the wider community. Much is still in flux here, which is preventing most companies from taking a closer look at the possibilities offered by quantum technology.

Introducing fundamental quantum computing concepts through use-case-driven scenarios in the form of a business simulation game within a fictitious company has proven to be a promising approach for an effective continuing education and training strategy. This approach enables participants to be addressed precisely at their current level of understanding and with their specific questions in mind.

Insights gained from the workshop conducted during the IT-Tage suggest that a single-day format tends to be too limited in duration, especially for participants with little or no prior knowledge in quantum technologies. Future workshops would benefit from an even stronger emphasis on practical, enterprise-relevant applications — for example, by incorporating a broader range of illustrative examples in the introductory sessions.

One of the key challenges of the workshop concept lies in the wide range of expertise required from the facilitators. In the implementation described here, this was effectively managed through the complementary subject-matter specializations of the instructors, which collectively ensured the necessary breadth of knowledge.

Future research should explore how the workshop concept can be adapted to specific industry domains, such as finance, logistics, or manufacturing, to enhance practical relevance for diverse corporate contexts. In parallel, studies should investigate how quantum computing education can be effectively integrated into existing corporate training and innovation programs, ensuring alignment with broader digital transformation strategies. The existing skills framework could also be supplemented by defining role-specific quantum computing skills for IT professionals - such as developers, architects or decision-makers - in order to support targeted and sustainable personnel development. Furthermore, tools need to be developed to assess learning success, such as quizzes. Another possibility is the development of standardized certifications in the field of vocational training by an independent consortium, as is already common practice in many areas of IT.

In summary, the user-centered introduction and transfer of quantum computing knowledge through interactive workshops represents a highly promising approach. By directly involving participants, this format facilitates a more targeted and responsive engagement with their individual needs and expectations, thereby enhancing the effectiveness and sustainability of the learning experience.

\backmatter


\bmhead{Acknowledgements}
We gratefully acknowledge the organizers of the IT Days for providing the opportunity to conduct a quantum computing workshop as part of the conference program. The authors would also thank the unknown reviewers for their valuable hints.

\section*{Declarations}

\begin{itemize}
\item Funding: None

\item Conflict of interest/Competing interests: 
The authors have no relevant financial or non-financial interests to disclose.

\item Ethics approval and consent to participate:
Not applicable. 

\item Consent for publication:
Not applicable.

\item Data availability:
The data that support the findings of this study are available upon reasonable request.

\item Materials availability:
Not applicable.

\item Code availability:
Not applicable.

\item Author contribution:
The authors contributed equally to the paper. All authors read and approved the final manuscript.
\end{itemize}

\bibliography{QuantumWorkshop}

@ARTICLE{Greinert_2023,
title={Towards a quantum ready workforce: the updated {E}uropean {C}ompetence {F}ramework for {Q}uantum {T}echnologies},
year={2023},
author={Franziska Greinert and Rainer Müller and Simon Goorney and Jacob Sherson and Malte Ubben},
doi={10.3389/frqst.2023.1225733},pmid={null},pmcid={null},mag_id={4384339208},
journal={Frontiers in Quantum Science and Technology}
}

@ARTICLE{Bonilla-Licea_2024,
title={A {P}rimer on {Q}uantum {M}echanics for {E}lectrical {E}ngineers},
year={2024},
author={M. Bonilla-Licea and Daniel Bonilla-Licea and Moisés Bonilla-Estrada and A. Soria-López and Juan Carlos Martínez-García},
doi={10.1109/access.2024.3444822},pmid={null},pmcid={null},mag_id={null},journal={IEEE Access}}

@ARTICLE{Aparicio-Morales_2024,
title={Supply and {D}emand in the {T}raining of {Q}uantum {S}oftware {E}ngineering {W}orkforce},
year={2024},
author={Álvaro M. Aparicio-Morales and E. Moguel and José Garcia-Alonso and Alejandro Fernández and Luis Mariano Bibbo and J. M. Murillo},
doi={10.36561/ing.27.16},pmid={null},pmcid={null},mag_id={null},
journal={Memoria Investigaciones en Ingeniería},
}

@ARTICLE{Kaur_2022,
title={Defining the quantum workforce landscape: a review of global quantum education initiatives},
year={2022},
author={Maninder Kaur and Araceli Venegas‐Gomez},
doi={10.1117/1.oe.61.8.081806},pmid={null},pmcid={null},mag_id={4280583859},
journal={Optical Engineering}
}

@ARTICLE{Venegas‐Gomez_2020,
title={The Quantum {E}cosystem and {I}ts {F}uture {W}orkforce},
year={2020},
author={Araceli Venegas‐Gomez},
doi={10.1002/phvs.202000044},pmid={null},pmcid={null},mag_id={3096277807},journal={PhotonicsViews}
}

@ARTICLE{Gerke_2022,
title={Requirements for future quantum workforce – a {D}elphi study},
year={2022},
author={F Gerke and R Müller and P Bitzenbauer and M Ubben and K-A Weber},
doi={10.1088/1742-6596/2297/1/012017},pmid={null},pmcid={null},mag_id={4287986102},
journal={Journal of physics}
}

@ARTICLE{Tenjo-Patino_2024,
title={Quantum {C}omputing {E}ducation in {L}atin {A}merica: {E}xperiences and {S}trategies},
year={2024},
author={Laura Tenjo-Patino and Cristian E. Bello and A. M. Cañola},
doi={https://doi.org/10.48550/arXiv.2410.18307},
}

@ARTICLE{Purohit_2024,
title={Qureka! {B}ox -- An {ENSAR} methodology based tool for understanding quantum computing concepts}
,year={2024}
,author={Abhishek Purohit and Jose Jorge Christen and Richard Kienhoefer and Simon Armstrong and Maninder Kaur and Araceli Venegas-Gomez Qureca Ltd. and Glasgow and Scotland and United Kingdom. and James Watt School Of Engineering and Quantum NanoPhotonics Group and U. Glasgow and Qureca Spain S.L. and Castelldefels and Barcelona and Spain and Computer Engineering Department and Computer Engineering Division and University Of Monterrey and Nuevo León and Mexico}
,doi={10.48550/arXiv.2410.21219},
}

@ARTICLE{Greinert_2024,
title={Advancing quantum technology workforce: industry insights into qualification and training needs},
year={2024},
author={Franziska Greinert and M. Ubben and Ismet N. Dogan and Dagmar Hilfert-Ruppell and Rainer Muller},
doi={10.1140/epjqt/s40507-024-00294-2},pmid={null},pmcid={null},mag_id={null},journal={EPJ Quantum Technology}}

@ARTICLE{Aiello_2021,
doi = {10.1088/2058-9565/abfa64},
url = {https://doi.org/10.1088/2058-9565/abfa64},
year = {2021},
month = {jun},
publisher = {IOP Publishing},
volume = {6},
number = {3},
pages = {030501},
title = {Achieving a quantum smart workforce},
journal = {Quantum Science and Technology},
author={Aiello, Clarice D and Awschalom, D D and Bernien, Hannes and Brower, Tina and Brown, Kenneth R and Brun, Todd A and Caram, Justin R and Chitambar, Eric and Di Felice, Rosa and Edmonds, Karina Montilla and Fox, Michael F J and Haas, Stephan and Holleitner, Alexander W and Hudson, Eric R and Hunt, Jeffrey H and Joynt, Robert and Koziol, Scott and Larsen, M and Lewandowski, H J and McClure, Doug T and Palsberg, Jens and Passante, Gina and Pudenz, Kristen L and Richardson, Christopher J K and Rosenberg, Jessica L and Ross, R S and Saffman, Mark and Singh, M and Steuerman, David W and Stark, Chad and Thijssen, Jos and Vamivakas, A Nick and Whitfield, James D and Zwickl, Benjamin M},

}

@ARTICLE{Greinert_2024a,
title={Extending the {E}uropean {C}ompetence {F}ramework for {Q}uantum {T}echnologies: new proficiency triangle and qualification profiles},
year={2024},
author={Franziska Greinert and Simon Goorney and Dagmar Hilfert-Ruppell and M. Ubben and Rainer Muller},
doi={10.1140/epjqt/s40507-024-00302-5},pmid={null},pmcid={null},mag_id={null},
journal={EPJ Quantum Technology}}

@ARTICLE{Chrisochoides_2024,
title={Developing a {F}ramework for {P}ersonalized {V}ideo-{B}ased {Q}uantum {I}nformation {S}cience {E}ducation},
year={2024},
author={Nikos Chrisochoides and N. Diawara and Michail Giannakos},
doi={10.1109/qce60285.2024.20457},pmid={null},pmcid={null},mag_id={null},
journal={International Conference on Quantum Computing and Engineering}
}

@Article{Hellstern_2024,
author={Hellstern, Gerhard
and Hettel, J{\"o}rg
and Just, Bettina},
title={Introducing quantum information and computation to a broader audience with {MOOCs} at {OpenHPI}},
journal={EPJ Quantum Technology},
year={2024},
month={Sep},
day={13},
volume={11},
number={1},
pages={59},
issn={2196-0763},
doi={10.1140/epjqt/s40507-024-00270-w},
url={https://doi.org/10.1140/epjqt/s40507-024-00270-w}
}

@BOOK{Hattie_2017_mindframes,
  title     = {10 {M}indframes for {V}isible {L}earning},
  author    = "Hattie, John and Zierer, Klaus",
  publisher = "Routledge",
  month     =  dec,
  year      =  2017,
  address   = "London, England",
  doi       = {10.4324/9781315206387}
}

@BOOK{Koenig_2020_Gruppen,
  title    = "Einf{\"u}hrung in die Gruppendynamik",
  author   = "K{\"o}nig, Oliver and Schattenhofer, Karl",
  publisher = "Carl Auer",
  month    =  jun,
  year     =  2020,
  address   = "Heidelberg, Germany",
  language = "de",
  note={{ISBN:} 978-3-8497-0344-8}
}

@BOOK{Aerssen_2018_Innovation,
  title     = "Das {große} Handbuch Innovation",
  editor    = "van Aerssen, Benno and Buchholz, Christian",
  publisher = "Vahlen, Franz",
  edition   =  1,
  month     =  aug,
  year      =  2018,
  address   = "Munich, Germany",
  language  = "de",
  doi       = {doi.org/10.15358/9783800656844}
}

@incollection{Wildt_2003_shift,
  title={"{T}he shift from teaching to learning" - {T}hesen zum {W}andel der {L}ernkultur in modularisierten {S}tudienstrukturen.},
  author={Wildt, Johannes},
  editor={Holger Ehlert},
  booktitle = {Qualitätssicherung und Studienreform},
  publisher = {Grupello-Verlag},
  pages={168--178},
  year={2004},
  address = {Düsseldorf},
  note={{ISBN:} 3-89978-023-X}
}

@BOOK{Herz_2000_Simulation,
  title     = "Simulation und Planspiel in {den} Sozialwissenschaften",
  author    = "Herz, Dietmar and Bl{\"a}tte, Andreas",
  year      = {2000},
  publisher = {Lit Verlag},
  address   = {Münster},
  series    = {Grundlegung und Methoden der politischen Wissenschaft},
  note      ={{ISBN:} 978-3-8258-4752-4}
}

@article{Dowling_2003,
  title={Quantum {T}echnology: {T}he {S}econd {Q}uantum {R}evolution},
  author={Dowling, Jonathan P. and Milburn, Gerard J.},
  journal={Phil. Trans. R. Soc. A},
  volume={361},
  pages={1655--1674},
  year={2003},
  doi={https://doi.org/10.1098/rsta.2003.1227}
}

@article{faisal2022business,
  title={Business {S}imulation {G}ames in {H}igher {E}ducation: A {S}ystematic {R}eview of {E}mpirical {R}esearch},
  author={Faisal, Nadia and Chadhar, Mehmood and Goriss-Hunter, Anitra and Stranieri, Andrew},
  journal={Human Behavior and Emerging Technologies},
  volume={2022},
  number={1},
  pages={1578791},
  year={2022},
  publisher={Wiley Online Library},
  doi={https://doi.org/10.1155/2022/1578791}
}
\end{document}